\documentclass{aa}

\usepackage{amsmath}
\usepackage{epsfig}
\usepackage{graphicx}
\usepackage{lscape}
\usepackage{subfig}
\usepackage{natbib}
\usepackage{txfonts}

\usepackage{amssymb}
\usepackage{color}
\usepackage{url}

\newcommand {\bc}{\begin {center}}
\newcommand {\ec}{\end {center}}
\newcommand {\be}{\begin {equation}}
\newcommand {\ee}{\end {equation}}
\newcommand {\beq}{\begin {eqnarray}}
\newcommand {\eeq}{\end {eqnarray}}

\def\flux{erg s$^{-1}$ cm$^{-2}$}
\def\lum{erg s$^{-1}$}

\def\igr{IGR~J19294+1816}

\begin{document}

   \title{Study of the X-ray pulsar IGR\,J19294+1816 with NuSTAR: detection of cyclotron line and transition to accretion from the cold disc}

   \author{Sergey~S.~Tsygankov \inst{1,2}
          \and  Victor~Doroshenko \inst{3}
          \and  Alexander~A.~Mushtukov \inst{4,2,5}
          \and  Alexander~A.~Lutovinov \inst{2}
          \and  Juri~Poutanen \inst{1,2}
          }

   \institute{Department of Physics and Astronomy, FI-20014 University of Turku, Finland;  
              \email{sergey.tsygankov@utu.fi}
       \and
             Space Research Institute of the Russian Academy of Sciences, Profsoyuznaya Str. 84/32, Moscow 117997, Russia
       \and
              Institut f\"ur Astronomie und Astrophysik, Universit\"at T\"ubingen, Sand 1, D-72076 T\"ubingen, Germany
       \and
            Leiden Observatory, Leiden University, NL-2300RA Leiden, the Netherlands
       \and
           Pulkovo Observatory of the Russian Academy of Sciences, Saint Petersburg 196140, Russia
          }
   \titlerunning{Study of X-ray pulsar IGR\,J19294+1816}
   \authorrunning{S.~S.~Tsygankov et al. }
   \date{Received 06.07.2018; accepted 19.11.2018}


  \abstract
  {In the work we present the results of
  two deep broad-band observations of the poorly studied X-ray pulsar
  \igr\ obtained with the {\it NuSTAR} observatory. The source was
  observed during Type I outburst and in the quiescent state. In the
  bright state a cyclotron absorption line in the energy spectrum was
  discovered at $E_{\rm cyc}=42.8\pm0.7$~keV. Spectral and timing
  analysis prove the ongoing accretion also during the quiescent state
  of the source. Based on the long-term flux evolution, particularly
  on the transition of the source to the bright quiescent state with
  luminosity around $10^{35}$ \lum, we concluded that \igr\ switched
  to the accretion from the `cold' accretion disc between Type I
  outbursts. We also report the updated orbital period of the system.
}


   \maketitle

%

\section{Introduction}
\label{intro}

Timing and spectral properties of radiation generated by accreting
objects carry information about physical and geometrical properties of the
systems. In the case of highly magnetized neutron stars (X-ray pulsars; XRPs)
detailed analysis of the emission in different luminosity states allows us to
investigate physical processes both very close to the neutron star (NS)
and at the boundary between accretion disc and the magnetosphere. Moreover,
observational appearance of these physical processes may drastically depend on
the properties of the particular XRP (pulse period, magnetic field strength,
etc.), reflecting different regimes of matter interaction with magnetic field
and radiation.

Accurate knowledge of the magnetic field strength is crucial for
application of physical models describing such interaction. 
Here we use high-quality {\it NuSTAR} data in order to accurately
measure the magnetic field of the poorly studied transient XRP \igr. This
information is further used to fill the gap in our knowledge of how pulsars
with intermediate spin periods interact with the accretion disc. In
particular, we aimed to verify our earlier prediction that there is a critical spin
period (around 10 s for standard magnetic field strength)
dividing all XRPs into two families: (i) short-spin pulsars able to
switch to the `propeller' regime at the final stages of their outbursts, and
(ii) long-spin ones continuing to accrete stably from the `cold' accretion
disc \citep{2017A&A...608A..17T}. \igr\ with $\sim$12.4~s spin period
\citep{2009ATel.1998....1R,2009ATel.2002....1S} is an excellent candidate to
fill the gap between these two families of XRPs.

\igr\ was discovered by the {\it INTEGRAL} observatory on 2009 March 27
\citep{2009ATel.1997....1T}. Analysis of the archival {\it Swift}/XRT data of
this region revealed relatively bright source showing evidence of pulsations at
$\sim$12.4~s \citep{2009ATel.1998....1R}. The pulsar nature of the source was
confirmed later by \cite{2009ATel.2002....1S}. Long-term flux variability with
period of about 117.2 d was discovered by the {\it Swift}/BAT monitor and was
associated with orbital modulation \citep{2009ATel.2008....1C}.
Based on the transient behaviour of \igr\ and its position on the Corbet
diagram an assumption about a Be/XRP nature of the source was made. The NIR
spectroscopy presented by \cite{2018MNRAS.476.2110R} directly confirmed this
hypothesis resulting in the identification of the optical companion in the
system with a B1Ve star located at the far edge of the Perseus arm at a
distance of $d=11\pm1$~kpc.

The relative faintness of \igr\ did not allow to make any conclusions regarding
the physical properties of the NS in the system up to date. Only tentative
detection of the cyclotron absorption line at $\sim$35.5~keV was reported based
on the deviation of the source spectrum from a power law at higher energies
using the RXTE data \citep{2017ApJ...848..124R}. In this work we use
long-term {\it Swift}/XRT monitoring observations to characterise accretion
regimes in the system as well as two dedicated deep {\it NuSTAR} observations
to describe properties of the NS in the hard X-ray range for the first time.

\section{Data analysis}

This work is based on the data from the XRT telescope
\citep{2005SSRv..120..165B} onboard the {\it Neil Gehrels Swift Observatory}
\citep{2004ApJ...611.1005G} obtained during monitoring programs performed
during and after Type I outbursts in 2010 and 2017--2018, as well as two {\it
NuSTAR} \citep{2013ApJ...770..103H} observations with the first one done nearly
simultaneous with {\it Swift}/XRT in March 2018.

\subsection{{\it Swift}/XRT data}

High sensitivity and flexibility of the {\it Swift}/XRT telescope allow us to
carry out long-term monitoring programs probing source flux evolution in a very
broad range. Particularly, it permitted us to investigate the transition of
\igr\ from the outburst to the quiescent state. Because of the low count rates all
XRT observations were performed in the Photon Counting (PC) mode and
automatically reduced using the online tools \citep{2009MNRAS.397.1177E}  provided by the UK
Swift Science Data Centre. \footnote{\url{http://www.swift.ac.uk/user_objects/}}

Data sample consists of observations performed after Type I outbursts occurred
in the end of 2010 and 2017. The corresponding light curves are shown in the
top panel of Fig. \ref{fig:lc}. Luminosity was calculated from the bolometric
(see below) and absorption corrected flux determined based on the results of
spectral fitting in {\sc xspec} package assuming absorbed power law model and
distance to the source $d=11$~kpc \citep{2018MNRAS.476.2110R}. Taking into
account low count statistics we binned the spectra in the 0.5--10 keV range to have
at least 1 count per energy bin and fitted them using W-statistic
\citep{1979ApJ...230..274W}.\footnote{see {\sc xspec} manual;
\url{https://heasarc.gsfc.nasa.gov/xanadu/}
\url{xspec/manual/XSappendixStatistics.html}}

In order to convert the observed 0.5--10 keV flux into the total X-ray
luminosity and, correspondingly, mass accretion rate, we estimated the
bolometric correction factors using two broad-band {\it NuSTAR} observations
performed in the quiescent and outburst states. Flux ratio between these
two states was more than 50. As will be discussed later, the broad-band
spectrum of the source depends slightly on its intensity, that is reflected in
the flux-dependent bolometric correction factors. Particularly, in the
quiescent state with the unabsorbed 0.5--10 keV flux around $F_{\rm 0.5-10
keV}\sim4\times10^{-12}$ \flux\ the bolometric correction (defined as the ratio of
fluxes in the 0.5--100 to 0.5--10 keV ranges) was $K_{\rm bol}\sim1.8$,
whereas during the second observation ($F_{\rm 0.5-10 keV}\sim1.1\times10^{-10}$
\flux) it increased up to $\sim2.5$.  Assuming a linear dependence of the
bolometric correction on the logarithm of flux in the 0.5--10 keV band a simple
equation can be obtained for $K_{\rm bol}=7.5 + 0.5 {\rm log}(F_{\rm
0.5-10 keV})$. In the following analysis we apply this correction to all XRT
observations and refer to the bolometrically corrected fluxes and luminosities,
unless stated otherwise.

\begin{figure}
\centering
\includegraphics[width=0.98\columnwidth, bb=20 155 565 705]{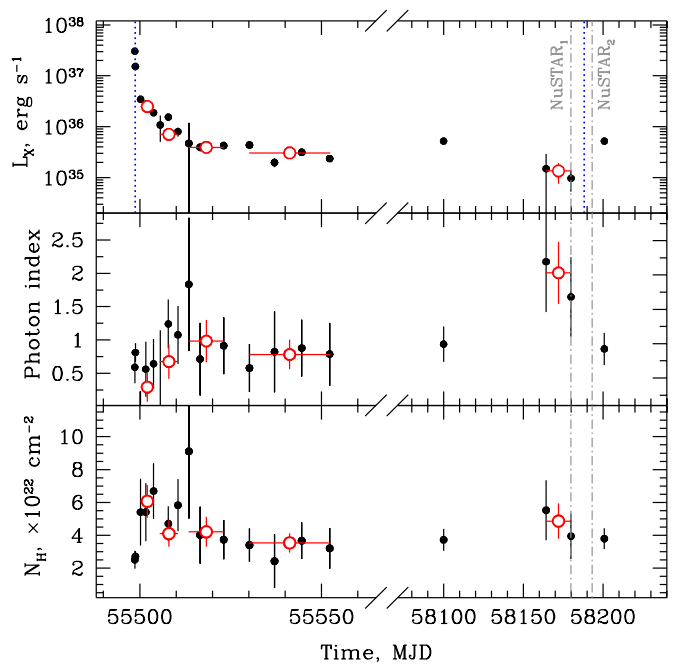}
\caption{{\it Top}: Bolometrically corrected light curve of
  \igr\ obtained with the {\it Swift}/XRT telescope in 2010 (left
  side) and 2017--2018 (right side) assuming distance to the source 11
  kpc. Blue dotted lines show moments of periastron passages (see
  Sect.~\ref{sec:orb}). {\it Middle and bottom}: The corresponding
  evolution of the power law photon index and absorption value
  assuming an absorbed power law model. Black points correspond to
  individual XRT observations, whereas red open circles represent
  parameters obtained from the averaging of a few nearby observations
  with low count statistics. Vertical dash-dotted lines show moments
  of the {\it NuSTAR} observations.}\label{fig:lc}
\end{figure}

\subsection{{\it NuSTAR} data}

The {\it NuSTAR} instruments include two co-aligned identical X-ray
telescope systems allowing to focus X-ray photons in a wide energy range from 3
to 79 keV \citep{2013ApJ...770..103H}. Thanks to the unprecedented sensitivity
in hard X-rays, {\it NuSTAR} is ideally suited for the broadband spectroscopy
of different types of objects, including XRPs, and searching for the
cyclotron lines in their spectra.

\igr\ has been observed with {\it NuSTAR} twice in March 2018 (see Table~\ref{tab1}). First
observation (ObsID 90401306002) was performed on March 3, in the very end of
the orbital cycle when the source was in the lowest state ever observed. Second
observation (ObsID 90401306004) occurred two weeks later, on March 16, when the
source entered another regular Type I outburst. In this state \igr\ was about
50 times brighter in comparison with the first observation.

\begin{table}
	\caption{The {\it NuSTAR} observations of \igr. }
	\label{tab1}
	\centering
	\begin{tabular}{lcccc}
		\hline
		ObsID & $T_{\rm start}$, & $T_{\rm stop}$, & Exp., & Net count\\
                      &  MJD         &   MJD       &  ks       &  rate, cts s$^{-1}$ \\
		\hline
		90401306002 & 58179.91 & 58180.84 & 40 & 0.06 \\  
		90401306004 & 58193.13 & 58194.06 & 40 & 2.03 \\       
		\hline
	\end{tabular}
\end{table}

The raw observational data were processed following the standard data reduction
procedures described in {\it NuSTAR} user guide and the standard {\it NuSTAR}
Data Analysis Software ({\sc nustardas}) v1.6.0 provided under {\sc heasoft
v6.24} with the CALDB version 20180419.

The source spectra were extracted from the source-centered circular region with
radius of 47\arcsec\ using the {\sc nuproducts} routine. The extraction radius
was chosen to optimize the signal to noise ratio above 30\,keV. The background
was extracted from a source-free circular region with radius of 165\arcsec\ in
the corner of the field of view.

\section{Results}
\label{sec:res}

The bolometrically and absorption corrected lightcurve of \igr\ obtained with
the {\it Swift}/XRT telescope during the monitoring programs in 2010 and
2017--2018 is shown on the top panel of Fig.~\ref{fig:lc}. The overall behaviour
of the source can be divided into two main states: (i) quiescent state with
luminosity around $10^{35}$ \lum, and (ii) outbursts associated with the
periastron passage (Type I outbursts) with peak luminosity reaching
$\sim10^{37}$ \lum. Middle and bottom panels of the figure show the
corresponding evolution of the photon index and absorption value measured in
the 0.5--10 keV energy band assuming absorbed power law model ({\sc phabs
  $\times$ pow} in {\sc xspec}). As can be seen, this simple spectral model fits the data in XRT range at all observed flux levels. Although some spectral variability is observed, the transition to the thermal spectrum, expected in the case of cooling NS surface \citep[see e.g.,][]{2016MNRAS.463L..46W}, is never observed in \igr, strongly indicating the continuation of accretion in the quiescent state.

Similar behaviour with transition of the source to the stable accretion between
Type I outbursts was recently discovered in another Be/XRP GRO J1008--57 and
interpreted as accretion from the cold disc \citep{2017A&A...608A..17T}. To
study this process in more details deep broad-band observations were requested
in these two states of \igr. In spite of only 13 days gap between observations {\it
NuSTAR} found the source in completely different states with luminosities
$L_{\rm X}=6.7\times10^{34}$ \lum\ (ObsID 90401306002; low state) and
$L_{\rm X}=3.4\times10^{36}$ \lum\ (ObsID 90401306004; high state).
The light curves of the source obtained from the {\it NuSTAR} data in full energy range do not reveal any strong variability. The source flux remains stable within a factor $2-3$ in both observations.
  
\subsection{Pulse profile and pulsed fraction}

\begin{figure}
\centering
\includegraphics[width=0.98\columnwidth, bb=20 160 555 680]{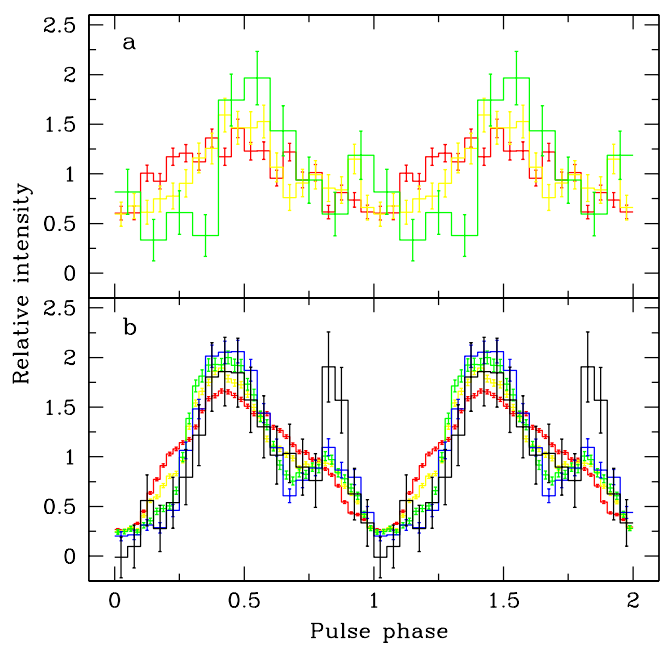}
\caption{Pulse profiles of \igr\ obtained with {\it NuSTAR} in
  different energy bands during (a) low and (b) high states. Different
  colours correspond to different energy bands: 3--10 keV (red),
  10--18 keV (yellow), 18--27 keV (green), 27--37 keV (blue), 37--50
  keV (black).}\label{fig:pprof}
\end{figure}

No binary orbital parameters except orbital period are known for \igr. Therefore pulsations
were searched in the light curves with only barycentric correction applied and
resulting periods might be biased due to orbital motion of the source.
Pulsations were detected with high significance in both states. The obtained
count statistics of {\it NuSTAR} data allowed us to reconstruct pulse profiles in several energy
bands from 3 to 50 keV for each observation using spin periods $P_1 =
12.4832(2)$~s and $P_2 = 12.4781(1)$~s for the low and high states,
respectively (see Fig.~\ref{fig:pprof}). Uncertainties for the pulse periods
were determined from large number of simulated light curves following
procedures described in \cite{2013AstL...39..375B}.

The shape of the pulse profile is very similar in different states. At the same time it demonstrates
a clear dependence on energy. Below $\sim10$ keV the profile has broad single
peak. With the increase of energy the profile
structure becomes more complicated. Particularly, it becomes double-peaked with
peaks separated by 0.5 in pulsar phase.

\begin{figure}
\centering
\includegraphics[width=0.98\columnwidth, bb=30 265 550 680]{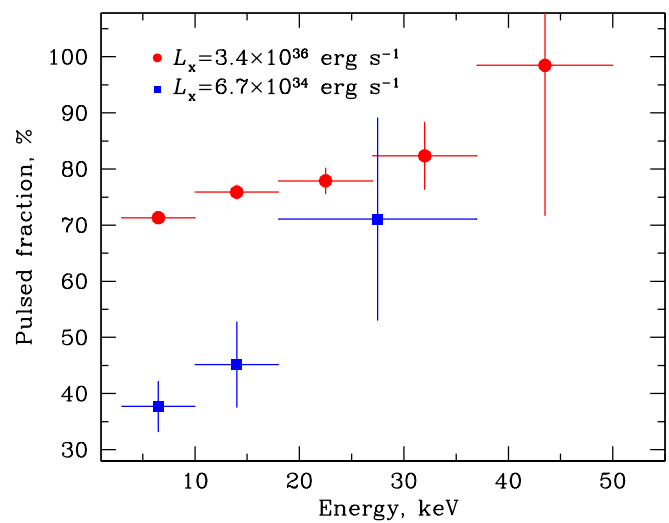}
\caption{Dependence of the pulsed fraction of \igr\ on energy as seen by {\it NuSTAR} in
  the low (blue squares) and the high (red circles)  states.}\label{fig:pfrac}
\end{figure}

The pulsed
fraction\footnote{$\mathrm{PF}=(F_\mathrm{max}-F_\mathrm{min})/(F_\mathrm{max}+F
_\mathrm{min})$, where $F_\mathrm{max}$ and $F_\mathrm{min}$ are maximum and
minimum fluxes in the pulse profile, respectively.} as a function of energy is
shown in Fig.~\ref{fig:pfrac}. For both states the increase of the pulsed
fraction towards higher energies is observed, that is typical behaviour for the
majority of XRPs \citep{2009AstL...35..433L}. Note that during the low
state the pulsed fraction was significantly lower. Similar drop of the pulsed
fraction in the low state was found in GRO~J1008--57 reinforcing the similarity between these two sources.

\subsection{Phase averaged spectral analysis}

Discovery of the cyclotron resonance scattering feature at $\sim36$\,keV in the
spectrum of the source has been claimed by \cite{2017ApJ...848..124R} based on
the observed deviation of the {RXTE}/PCA continuum from a power law above
$\sim30$\,keV. The choice of the continuum model by the authors appears,
however, to be extremely odd, because X-ray spectra of all known XRPs exhibit a
cutoff above $\sim15-20$\,keV \citep[see e.g.,][]{1989PASJ...41....1N,2005AstL...31..729F}. 
The cutoff is actually
expected for a spectrum produced through comptonization in hot emission region
\citep[e.g.][]{1980A&A....86..121S,1985ApJ...299..138M}. It is extremely
likely, therefore, that the cyclotron line included in the model in practice
just mimicked the cutoff, and thus is not real.

To verify this possibility, we re-analyzed the {RXTE} observation
95438-01-01-00 (MJD~55499.46) where the line appeared to be most significant.
Indeed, we found that the broadband {RXTE}/PCA continuum is perfectly well
described with an absorbed cut-off power law model with no additional features
required by the fit. Alternatively, it can be described with one of the
comptonization models such as {\sc comptt} or {\sc nthcomp} with
$\chi^2/{\rm dof}\sim0.8$ (in all cases the fluorescence iron line was also included in
the fit with the energy fixed at 6.4\,keV and width at 0.1 keV). On the other
hand, we verified that it is indeed possible to model the spectrum with a
combination of a power law and {\sc cyclabs} model, however, not only the fit
quality is considerably worse in this case but also the line itself gets
unreasonably deep. We conclude, therefore, that the claim of the cyclotron line
discovery by \cite{2017ApJ...848..124R} is unsupported by the data used by the
authors, which has insufficient statistics.

\begin{figure}
\centering
\includegraphics[width=\columnwidth]{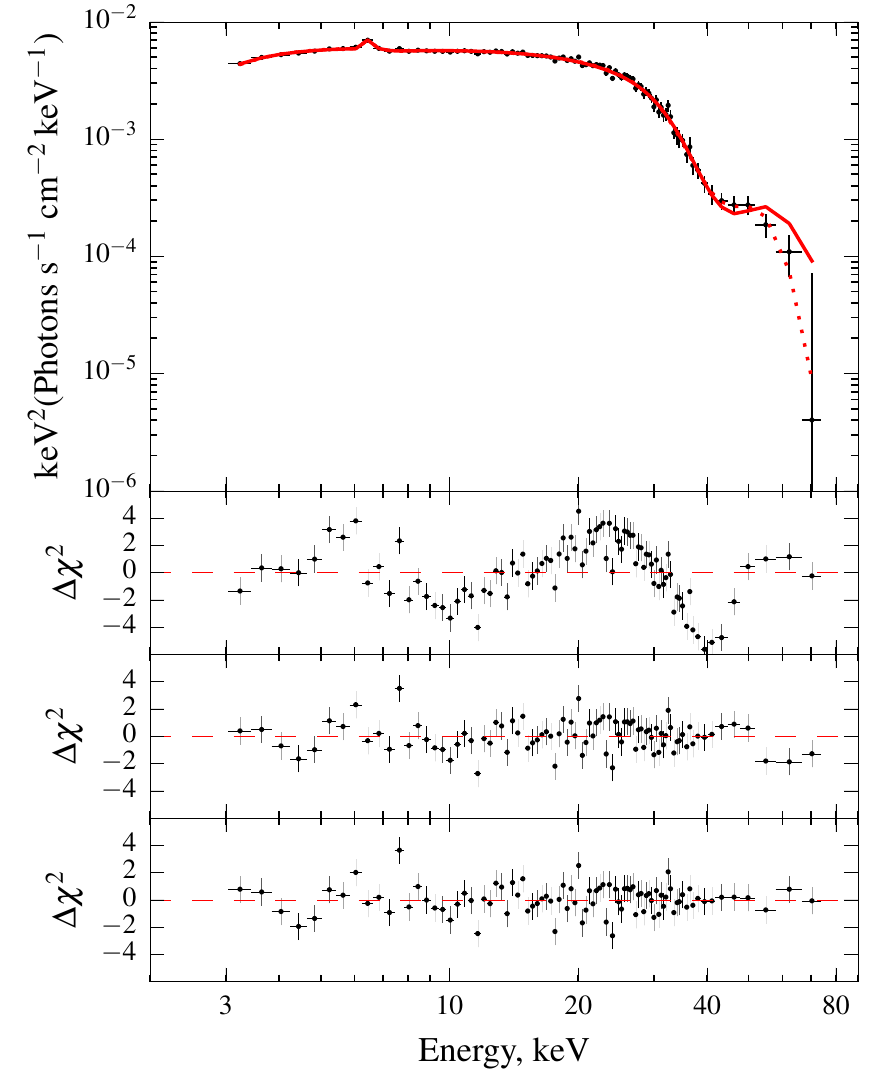}
\caption{Spectrum of \igr\ obtained with the {\it NuSTAR}
telescopes during the high state. The data from the two {\it
NuSTAR} units are added together for plotting (but not for actual fit). The
best-fitting residuals for models (top to bottom) without inclusion of the
absorption feature, with fundamental only (at $\sim42.8$\,keV; shown with the solid line in the top panel), and 
including also the harmonics (at $\sim85$\,keV, dotted line in the top panel) are also shown.
}\label{fig:spe}
\end{figure}

On the other hand, {\it NuSTAR} data provide much better statistics
and allow us to conduct a much more sensitive search for the possible
cyclotron lines in the source spectrum. Similarly to RXTE, {\it NuSTAR}
broadband spectrum can be well described with several continuum models.
However, residuals around $\sim40$\,keV in absorption are immediately apparent in phase-averaged spectrum from 
observation 90401306004 (i.e. when the source was in bright state)  irrespective
on the continuum model used. The fit can be greatly improved by inclusion of
a gaussian absorption line in the model. The width of the line depends slightly
on the continuum model, but the feature is always significantly detected at the same energy.
For the {\sc nthcomp} continuum model, an absorption line with energy 42.9(1)\,keV, width of 6.9(5)\,keV and optical depth at line centre of $\tau=1.3(2)$ improves the fit from $\chi^2/{\rm dof}=1132/720$ to $\chi^2/{\rm dof}=766/717$, which
corresponds to probability of chance improvement of $\sim3\times10^{-79}$ according to MLR test
\citep{2002ApJ...571..545P}.

The feature is, therefore, highly significant. This is also the case with other continuum models, so we
conclude that while the report by \cite{2017ApJ...848..124R} is
erroneous, the source does have indeed a cyclotron line at
$\sim43$\,keV, which implies magnetic field of $\sim5\times10^{12}$\,G
assuming gravitational redshift $z=0.35$ for the typical NS
parameters.
Usage of the {\sc cyclabs} model instead of {\sc gabs}
results in slightly lower cyclotron energy about 40 keV. Such
discrepancy between these two models is associated with the definition of the latter model, and was found in
other studies
\citep[e.g.][]{2012MNRAS.421.2407T,2015MNRAS.454.2714M}.
There is also some evidence for the harmonic at double energy in the residuals
(see Fig.~\ref{fig:spe}), although its significance is low with false alarm
probability of $\sim2$\% (assuming the energy and width of the fundamental
fixed to be double that of the fundamental, and even lower otherwise). 

The counting statistics in the second observation, unfortunately, does not allow us 
to confidently detect the line, nor to constrain its parameters.
Constraining the luminosity-related changes of the cyclotron line energy is, however,
extremely interesting given the results reported previously for other sources \citep{2006MNRAS.371...19T,2007A&A...465L..25S,2012A&A...542L..28K}.
We attempted thus to estimate the line energy in the low state by formally including it in the
fit, and modelling the high and low-state spectra simultaneously with the absorption column
and cyclotron line width and depth linked between both observations (to reduce the number of free parameters).
The fit results are presented in Table~\ref{tab:spe}. We found the line energy to be consistent between
the observations, although uncertainties for the low-state observation are, unfortunately, rather large.
It is important to emphasise, that the low counting statistics precludes significant detection of the line
if only the second observation is considered.
A deeper observation in quiescence would thus be required to firmly detect the line and to constrain its
parameters.

\begin{figure}
\centering
\includegraphics[width=\columnwidth]{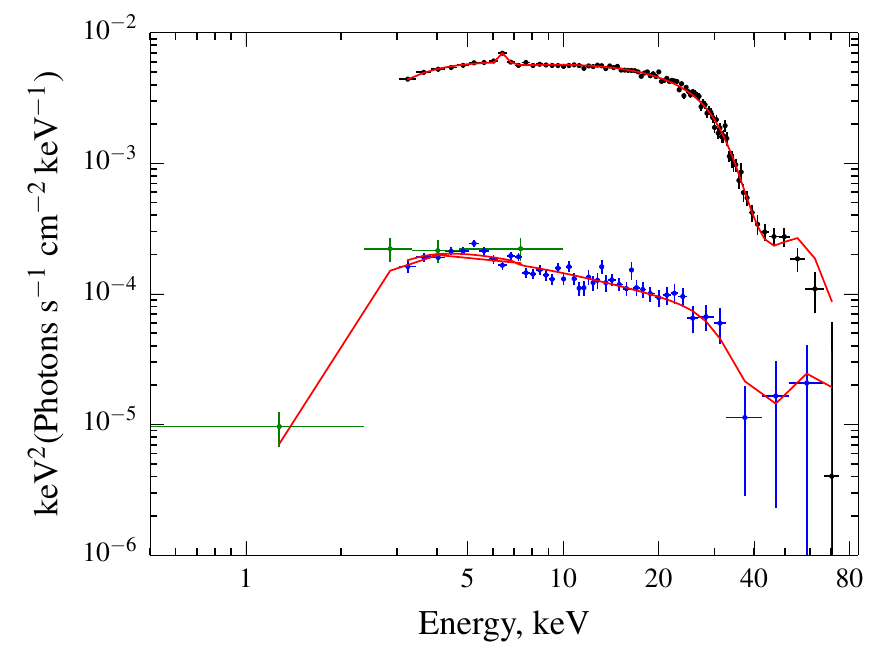}
\caption{Spectral energy distribution of \igr\ obtained with the {\it Swift}/XRT and {\it NuSTAR}
telescopes during the high (black dots) and low (blue and green dots) intensity states. The data from the two {\it
NuSTAR} units are added together for plotting (but not for actual fit). Solid lines represent best-fit models for each observation including fundamental line only (see text).
}\label{fig:spe2}
\end{figure}

\begin{table}
  \noindent
\caption{Best-fit parameters for a {\sc phabs*nthComp} model
  obtained for both observations. For the high state observation also the
  absorption feature modelled as {\sc gabs} and fluorescence iron line modelled as {\sc gaussian} are included in the fit.
  Neither feature was formally required for the low state spectrum.}
\label{tab:spe}
\begin{tabular}{cccc}
\hline
\hline
        Parameter &                     Units &           Low &                       High\\
\hline
      $N_{\rm H}$ & $10^{22}\text{ cm}^{-2}$ &   \multicolumn{2}{c}{$6.3\pm0.4$}\\
         $\Gamma$ &            --         & $1.68\pm0.03$ &        $1.443_{-0.006}^{+0.003}$\\
    $kT_{\rm e}$ &                      keV &  $15_{-4}^{+16}$  &          $7.0\pm0.2$\\
   $kT_{\rm bb}$ &                      keV &   \multicolumn{2}{c}{0.5$^{\rm fixed}$}\\
      $E_{\rm Fe}$ &     keV                 &             -- &     $6.40_{-0.04}^{+0.07}$\\
      $\sigma_{\rm Fe}$ &     keV             &            -- &     0.01$^{\rm fixed}$\\
      $N_{\rm Fe}$ &    $10^{-5}$ ph cm$^{-2}$ s$^{-1}$ &    -- &          $6.7\pm0.8$\\
        $E_{\rm cyc}$ &                       keV & $41.4_{-2.4}^{+3.0}$ &      $42.8\pm0.7$\\
   $\sigma_{\rm cyc}$ &                       keV &  \multicolumn{2}{c}{$6.8\pm0.5$}\\
        $\tau_{\rm cyc}$ &                    -- &   \multicolumn{2}{c}{$1.3\pm0.2$}\\
$\chi^2$/{dof} &                           &     \multicolumn{2}{c}{1.03 (850)} \\
\hline
\end{tabular}
\end{table}

\subsection{Phase resolved spectral analysis}

\begin{figure}
\centering
\includegraphics[width=\columnwidth]{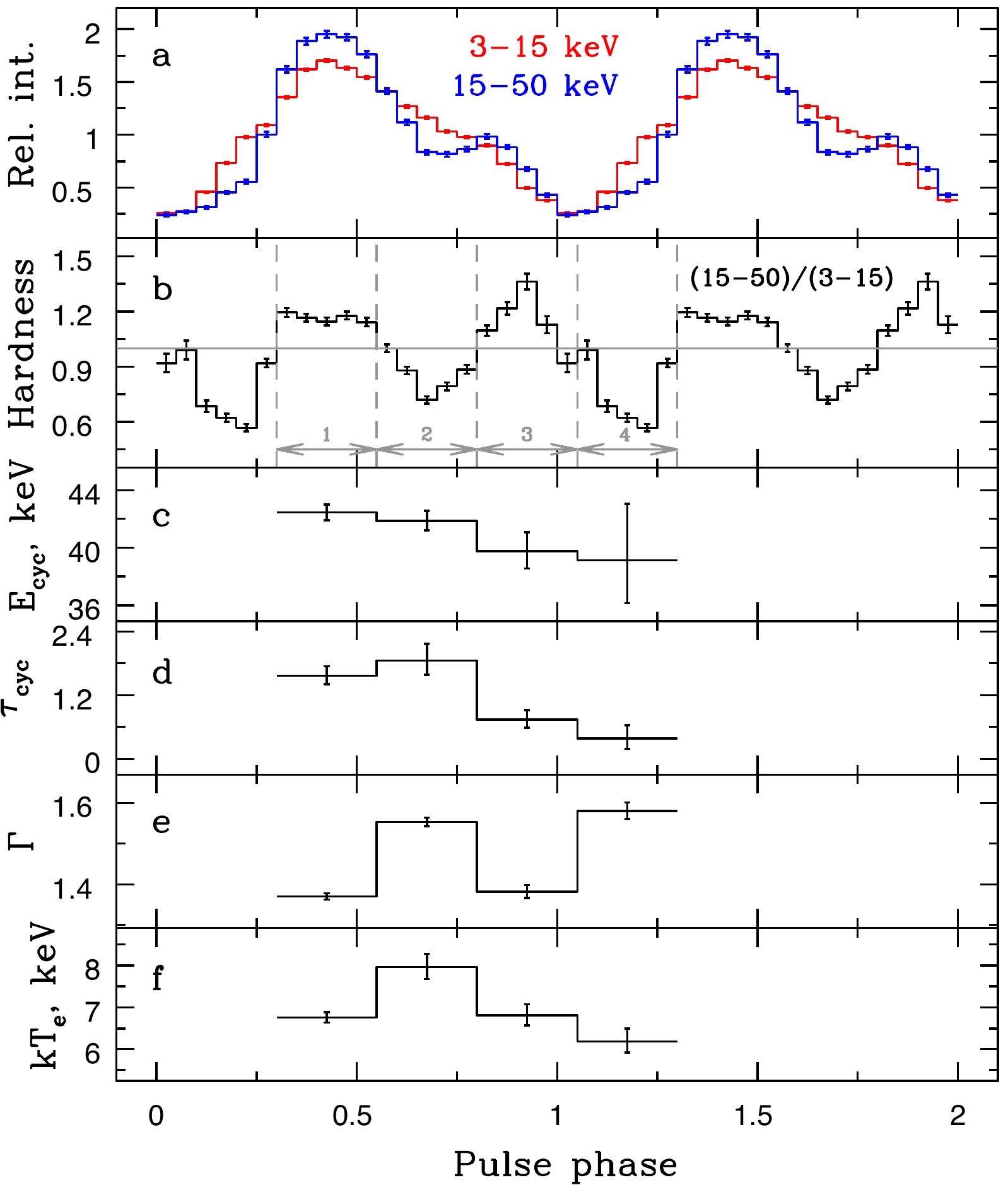}
\caption{Results of phase resolved spectroscopy of \igr\ obtained with the {\it NuSTAR}
telescope in the high intensity state. 
}\label{fig:phspec1}
\end{figure}

To investigate possible pulse phase dependence of spectral parameters in \igr\ we performed phase resolved spectral analysis in the bright state ({\it NuSTAR} ObsID 90401306004). Phase bins for individual spectra were chosen based on the hardness ratio over the pulse. As can be seen from the two top panels in Fig.~\ref{fig:phspec1}, there are four phase bins with significantly different values of hardness ratio, that can be a hint for significantly different spectral properties.

\begin{figure}
\centering
\includegraphics[width=\columnwidth]{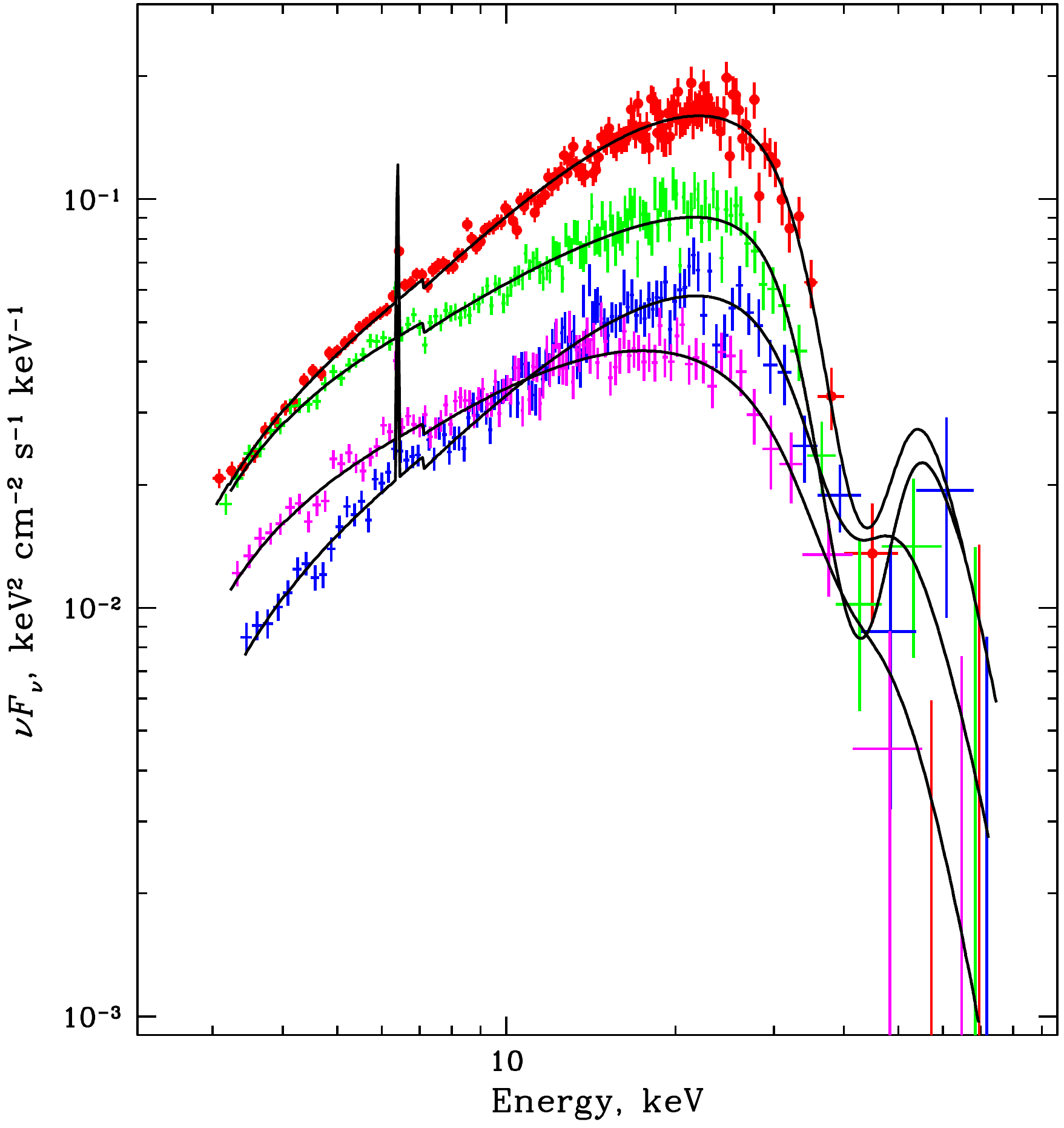}
\caption{Spectra of \igr\ obtained in four pulse phases shown in Fig.~\ref{fig:phspec1}. Phase bins 1, 2, 3 and 4 correspond to red, green, blue and black points, respectively. 
}\label{fig:phspec2}
\end{figure}

The obtained four spectra were fitted with the same model as in the case of
phase averaged spectrum, i.e. {\sc phabs*(gau+nthcomp*gabs)} in {\sc xspec}.
Absorption parameter $N_{\rm H}$ was free to vary and we did not find any
evolution of its value over the pulse period. Because of low count
statistics at high energies we froze the cyclotron line width at 6.8 keV determined from
phase-averaged spectrum (see Table~\ref{tab:spe}). Variability of the cyclotron line energy and optical depth
is shown in Fig.~\ref{fig:phspec1}c,d. The position
of the cyclotron line does not exhibit strong dependence on the pulse phase,
whereas depth of the line appears to vary significantly. In particular, the line is 
deepest at phases around 0.5--0.7 and is consistent with zero around phase 0.2.
Parameters of the continuum (both $\Gamma$ and $kT_{\rm e}$ were free to vary) are
significantly different at different phases, which is in line with observed hardness variations (see 
Fig.~\ref{fig:phspec1}e,f). For the illustrative purposes we show all four spectra
on one plot (see Fig.~\ref{fig:phspec2}).

\subsection{Orbital period}
\label{sec:orb}

\begin{figure}
    \centering
    \includegraphics[width=\columnwidth]{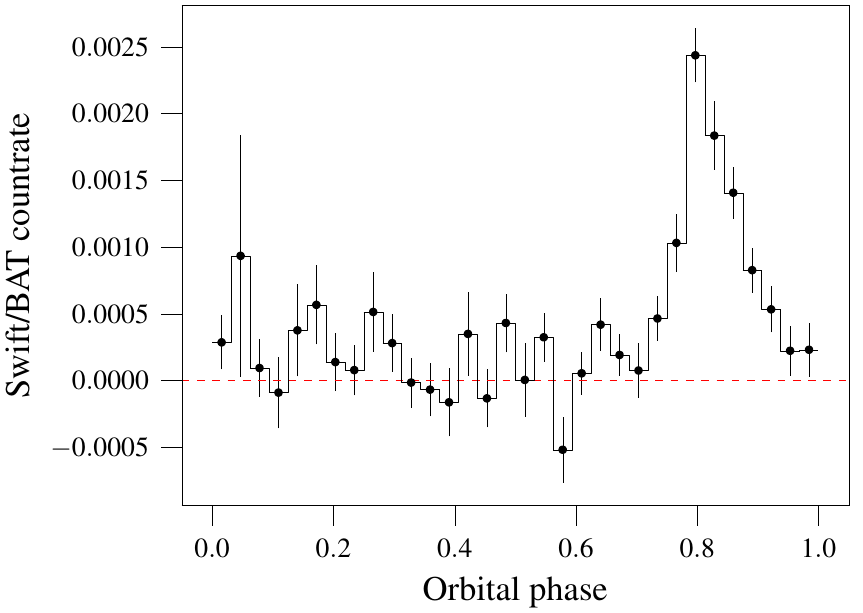}
    \includegraphics[width=0.9\columnwidth]{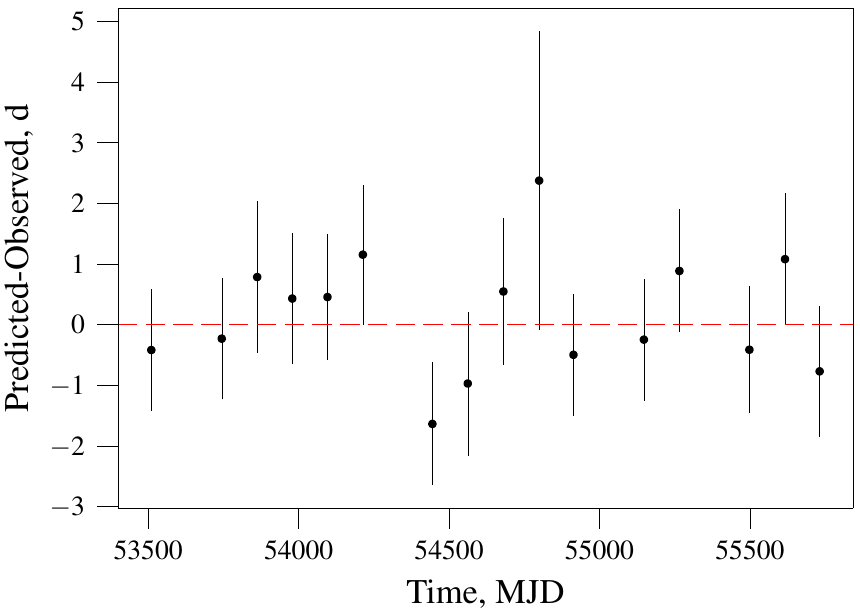}
    \caption{{\it Top}: Folded light curve of the source as observed by {\it Swift}/BAT. 
    The zero phase here is arbitrary to make the peak clearly visible.
    {\it Bottom}: the residuals to the linear  fit of the outburst peak times determined as described in the
      text. }
    \label{fig:porb}
\end{figure}

Orbital period of the source has been estimated by \cite{2009ATel.2008....1C}
at around 117\,d based on the variability of {\it Swift}/BAT light curve. A lot
of new data has been accumulated since 2009, so we exploited that to refine the
orbital period value. In particular, given the low duration and symmetric
profile of the outbursts, we determined the time corresponding to the peak of
each outburst in the light curve by fitting a gaussian in the light curve (for orbital cycles exhibiting any excess
around the peak). These times were then
used to determine the orbital period using the phase connection technique as
presented in Fig.~\ref{fig:porb}. The orbital period was found to be 116.93(5)~d,
with the outburst peak epoch of MJD 53510.9(5), which is in line with earlier
estimates \citep{2009ATel.2008....1C,2011A&A...531A..65B}.

\section{Discussion}
\label{sec:discus}

One of the main goals for our observational campaign of \igr\ was to investigate
the source behaviour during its transition from the state of intensive accretion
during Type I outbursts to the quiescent state between them.
Particularly, transition to the propeller regime, when accretion is prohibited
by the centrifugal barrier of the rotating NS magnetosphere, could be expected
in the case of relatively strong magnetic field \citep{1975A&A....39..185I}.

Observational appearance of transition to the propeller regime in classical
XRP is very clear and is expressed in a sharp drop of the source
luminosity by approximately two orders of magnitude, depending on the NS
magnetic field
\citep{1986ApJ...308..669S, 2016A&A...593A..16T, 2016MNRAS.457.1101T,2017ApJ...834..209L}.
A substantial softening of the energy spectrum is also expected from
accretion dominated power law to the NS cooling generated black body
\citep{2016A&A...593A..16T}. The threshold luminosity for the onset of the
propeller regime is determined by the equality of the co-rotation and
magnetospheric radii \citep[see e.g.,][]{2002ApJ...580..389C}:
\be \label{eq:Lprop}
L_{\rm prop}(R) \simeq \frac{GM\dot{M}_{\rm lim}}{R} 
\simeq 4 \times 10^{37} k^{7/2} B_{12}^2
P^{-7/3} M_{1.4}^{-2/3} R_6^5 \,\textrm{erg s$^{-1}$},
\ee
where $P$ is pulsar's rotational period in seconds, $B_{12}$ is 
NS magnetic field strength in units of $10^{12}$~G,  
$R_6$ is NS radius in units of $10^6$~cm and $M_{1.4}$ is the NS mass
in units of 1.4$M_\odot$. The factor $k$ relates the magnetospheric
radius in the case of disc accretion to the classical Alfv\'en radius
($R_{\rm m}=k R_{\rm A}$) and is usually assumed to be $k=0.5$, which appears justified both from theoretical and observational points of view \citep[see e.g., ][]{GL1979a,2014A&A...561A..96D, 2018A&A...610A..46C}.

However, recently it was shown that propeller effect can be observed only in
XRPs possessing relatively short spin period and/or strong magnetic field. In
the opposite case the cooling front, caused by thermal-viscous instability
\citep[see review by][]{2001NewAR..45..449L}, is able to reach inner radius of
the accretion disc resulting in the NS transition to the stable accretion from
the cold (low-ionization) disc before centrifugal barrier is able to fully cease accretion \citep{2017A&A...608A..17T}. First example of
such a behaviour was recently discovered in the Be/XRP GRO J1008--57, whose spin period is $\sim$94~s.

Luminosity corresponding to the transition of the whole accretion disc to the
cold state is determined by the inner radius of the disc and, therefore, by the strength of the NS magnetic field:
\be \label{eq:Lcold}
  L_{\rm cold}  \simeq  9\times 10^{33}\,k^{1.5}\,M_{1.4}^{0.28}\,R_6^{1.57}\,B_{12}^{0.86}~~~{\rm erg\,s^{-1}}.
\ee
Below this level, the temperature in the accretion disc is lower than 6500\,K at $R>R_{\rm m}$ \citep{2017A&A...608A..17T}.

As clearly seen from Eqs. (\ref{eq:Lprop}) and (\ref{eq:Lcold}) the behaviour of
the pulsar at low mass accretion rates is determined by the spin period of the
NS and its magnetic field. Thanks to the robust determination of the NS
magnetic field strength from the position of the cyclotron line we are able to apply physical
models of the accretion disc interaction with magnetosphere in \igr.
Substituting $P=12.4$~s and $B_{12}=5$ to Eqs. (\ref{eq:Lprop}) and
(\ref{eq:Lcold}) one can get $L_{\rm prop}=2.5\times10^{35}$~\lum\ and $L_{\rm
cold}=0.1\times10^{35}$~\lum\ for this source. We see that $L_{\rm
prop}\textgreater L_{\rm cold}$, which means that the source is expected to switch
to the propeller regime before the cooling wave will reach inner radius of the
accretion disc causing transition to the stable accretion.

However, the observed long-term light curve of the source, shown in
Fig.~\ref{fig:lc}, clearly demonstrates the transition of the pulsar into
stable state with luminosity around $2\times10^{35}$~\lum\ that is much higher
than expected from quiescent NS in XRPs emitting its thermal energy \citep[see
e.g.,][]{2017MNRAS.470..126T}. This is clear indication that \igr\ was able to
switch to the accretion from the cold disc before the centrifugal barrier
stopped the accretion.

The cold disc accretion scenario is also supported by the spectral and
timing analysis. First of all, a hard power-law-like spectral shape strongly
supports continuation of the accretion process. Just insignificant increase
of the photon index in the low state is observed (see middle panel of
Fig.~\ref{fig:lc}). Also the pulsed fraction drops significantly in this state,
possibly pointing to the increased emitting area. Worth noting that both
mentioned features of accretion from the cold disc were observed in
GRO~J1008--57 \citep[][]{2017A&A...608A..17T}.

The solution of this discrepancy between an expected and actual behaviour of
the source can be in the proximity of \igr\ parameters to the border between
pulsars expected to transit to the propeller regime and ones able to accrete
from the cold disc \citep[see Fig. 3 from][]{2017A&A...608A..17T}. 
Therefore, substantial simplifications contained in Eqs. (\ref{eq:Lprop})
and (\ref{eq:Lcold}) may become important for such sources. 
Some of the caveats have already been discussed in \cite{2017A&A...608A..17T}, others include neglection of the accretion torque and convection, which might affect inner disc temperature, and thus transitional luminosity. Detailed discussion of these factors is out of scope of this paper and will be published elsewhere.

Not only Eqs. (\ref{eq:Lprop}) and (\ref{eq:Lcold}) are very
approximate, but also the effective magnetosphere radius and thus
value of $k$ are rather uncertain \citep[see
  e.g.,][]{2015MNRAS.449.4288D,2017MNRAS.470.2799C,2017AstL...43..706F,2018A&A...617A.126B}. Finally,
transitional luminosity level also depends on the assumed distance,
which is also poorly known in most cases. It is essential, therefore,
to increase the sample of objects with intermediate spin periods
observed in quiescence, preferably including monitoring the
transition.

\section{Conclusion}

In the work we present the results of the long-term observational
campaign of poorly studied XRP \igr\ performed with the {\it
  Swift}/XRT telescope as well as two deep broad-band observations
obtained with the {\it NuSTAR} observatory.  It is shown that between
bright regular Type I outbursts with peak luminosity around $10^{37}$
\lum\ the source resides in a stable state with luminosity around
$10^{35}$ \lum. Spectral and timing properties of \igr\ point to the
ongoing accretion in this low state, which we interpreted as accretion
from the cold (low-ionization) accretion disc
\citep{2017A&A...608A..17T}.  In the bright state a cyclotron
absorption line in the energy spectrum was discovered at $E_{\rm
  cyc}=42.8\pm0.7$~keV allowing us to estimate the NS magnetic field strength
to around $5\times10^{12}$~G.  We were also able to substantially
 refine the orbital period of the system based on the long-term
\emph{Swift}/BAT light curve of the source.

\section*{Acknowledgements}
This work was supported by the grant 14.W03.31.0021 of the Ministry of
Science and Higher Education of the Russian Federation. This research was also supported by the 
Academy of Finland travel grants 309228 and 317552 (ST, JP) and by the Netherlands Organization for Scientific Research Veni Fellowship (AAM).
This work made use of data supplied by the UK Swift Science Data Centre at the University of Leicester.
We also express our thanks to the {\it NuSTAR} and {\it Swift}
teams for prompt scheduling and executing our observations.


\bibliographystyle{aa}
\bibliography{allbib}


\end{document}